\documentclass{osa-article}

\journal{osajournal}

\articletype{Research Article}

\usepackage{lineno}

\begin{document}

\title{Diffraction gratings based on multilayer silicon nitride waveguide with high upward efficiency and large effective length}

\author{
Wen-ling Li,\authormark{1,2,*}
Jing-wei Liu,\authormark{2}
Guo-an Cheng, \authormark{1}
Qing-zhong Huang, \authormark{3}
Rui-ting Zheng, \authormark{1}
Xiao-ling Wu, \authormark{1}
}

\address{\authormark{1}School of Nuclear Science and Technology, Beijing Normal University, China\\
\authormark{2}China Science Photon Chip (Haining) Technology Co., Ltd., China\\
\authormark{3}Wuhan National Laboratory for Optoelectronics, Huazhong University of Science and Technology, China}
\email{\authormark{*}wenlingli@mail.bnu.edu.cn} 



\begin{abstract}
Diffraction gratings with high upward diffraction efficiency and large effective length are required for chip-scale light detection and ranging. 
In this paper, we propose a diffraction grating based on a multilayer silicon nitride waveguide, which theoretically achieves an upward diffraction efficiency of 92$\%$, a near-field effective length of 376 $\mu m$ and a far-field divergence angle of 0.105$^{\circ}$ at a wavelength of 850 nm. 
The diffraction grating has a high tolerance to process variations based on Monte Carlo Analysis.
When the conditions are $\pm$5$\%$ layer thickness variation, $\pm$50 nm lithographic variation and $\pm$20 nm wavelength drift, more than 71$\%$ of the grating samples have a diffraction efficiency higher than 80$\%$, and 100$\%$ of the samples have an effective length larger than 200 $\mu m$ (corresponding to a far-field divergence <0.2$ ^{\circ}$).
Furthermore, the near-field effective length of the grating with an upward diffraction efficiency above 90$\%$ can be adjusted from hundreds of microns to centimeters by changing the etching layer thickness and the grating duty cycle. This diffraction grating has potential application in optical sensing and imaging from visible to near-infrared wavelengths.
\end{abstract}

\section{Introduction}
Augmented reality and autonomous driving have shown a growing demand for highly integrated, solid-state and low-cost light detection and ranging (LIDAR) solutions. 
Therefore, chip-scale LIDAR has become a hot spot in research\cite{1,2,3}. 
Among the LIDAR chip designs, waveguide diffraction gratings are widely adopted as antennas to convert beams from waveguide mode to space radiation mode.

LIDAR requires a small beam divergence angle to distinguish a person or small obstacles at a long distance (typically, ~0.1$^{\circ}$ for a distance of 100 m), which means that the beam waist is close to the mm level at a wavelength of 1550 nm\cite{1,4}.
For end-fire chips, space optical elements are needed to enlarge the output spot size. 
While diffraction gratings have better beam control capability and can achieve large spot size on-chip\cite{4,5,6,7,8,9}. 
Furthermore, the grating diffraction angle is sensitive to changes in wavelength, which can be used to implement wavelength-dependent angle steering\cite{2,10} in addition to phased arrays.

However, there are still many challenges. 
An mm near-field effective length requires a sufficient low grating strength, which means a shallow etching depth of several nanometers for silicon waveguides with a high refractive index contrast between the core and cladding. That leads to difficulties in the etching process. 
Consequently, some researchers propose fabricating fully etched sidewall perturbations on the waveguide, which successfully achieves an effective length of 300-500 $\mu m$ at 1550 nm, corresponding to an FWHM far-field divergence angle of 0.18-0.4$^{\circ}$\cite{5,6}. 
The introduction of a silicon nitride etched layer with a lower refractive index in the grating can further increase the effective length to 1 mm, which corresponds to a far-field divergence angle of 0.089-0.1$^{\circ}$ at a wavelength of 1550 nm \cite{4,8}.

Silicon nitride waveguides are transparent from visible to near-infrared wavelengths and easily achieve longer grating effective lengths due to the lower refractive index contrast. 
Thus, low-cost chip-scale LIDAR working at the wavelength of 850-940 nm is possible by combing the emitter based on silicon nitride waveguide and receiver based on silicon avalanche photodiode or silicon photomultiplier\cite{11,12,13}. 
However, compared with that of silicon gratings, the upward emission of silicon nitride gratings is lower, and the crosstalk between adjacent gratings is higher.
To suppress the crosstalk among waveguide grating arrays, isolation structures\cite{5}, uneven pitch arrangements\cite{5,8,14} and slab diffraction gratings\cite{9,15,16} have been proposed. 
The slab diffraction grating scheme fundamentally avoids crosstalk among waveguide gratings and is experimentally demonstrated to achieve a large steering range\cite{15}.

To improve the grating unidirectional emission, conventional ideas such as introducing Bragg reflectors\cite{17} or metal mirrors\cite{18} in the cladding would increase the process complexity or be incompatible with the subsequent high-temperature processes.
An offset double layer silicon nitride waveguide grating structure does not require a reflector and is theoretically proven to have an upward diffraction efficiency of 93.2$\%$ and an effective length of 3 mm at a wavelength of 1550 nm\cite{7}. 
However, this structure still requires 2 steps of lithography and accurate overlay between masks, which is a complicated preparation process. 
The multilayer silicon nitride TriPleX waveguide has been demonstrated to have low propagation loss and high design flexibility\cite{19}, which provides the possibility of embedding a Bragg reflector or photonic bandgap structure into the core layer. 
To the best of our knowledge, this structure has not yet been reported in the field of diffraction gratings.

In this paper, a one-dimensional slab diffraction grating based on a multilayer silicon nitride waveguide at the wavelength of 850 nm is designed. 
By optimizing the grating efficiency based on photonic band-gap theory, high upward emission and a large near-field effective length can be achieved at the same time. 
Since no additional reflective layer is needed and only one lithography step is required on the diffractive region, the preparation process is simplified. Furthermore, we theoretically demonstrated that the grating performance has high tolerance to process variation and laser wavelength drift.

\section{Design and simulation}
The structure of the multilayer $\rm Si_{3}N_{4}$ waveguide grating is shown in Fig. 1. 
From bottom to top, the layers are the silicon substrate, the bottom cladding layer of $\rm SiO_{2}$, the first core layer of $\rm Si_{3}N_{4}$, the first gap layer of $\rm SiO_{2}$, the second core layer of $\rm Si_{3}N_{4}$, the second gap layer of $\rm SiO_{2}$, the third etched layer of $\rm Si_{3}N_{4}$, and the top cladding layer of $\rm SiO_{2}$. 
Each layer in the structure has a specific function during the diffraction process. 
The composite core layers (including the first core layer of $\rm Si_{3}N_{4}$, the first gap layers of $\rm SiO_{2}$ and the second core layer of $\rm Si_{3}N_{4}$) guide the light in waveguide mode from the input waveguide. 
While in the grating region, these core layers function as both a waveguide and a reflector, optimized to enhance the upward efficiency of the diffraction light. 
And the third etching layer of $\rm Si_{3}N_{4}$ acts as the diffracting layer, in which light diffracts in a specific direction by periodically scattering waveguide light.  
Additionally, the second gap layer of $\rm SiO_{2}$ below the diffracting layer acts as an etched buffer layer to increase the etched depth tolerance.


\begin{figure}[htbp]
\centering\includegraphics[width=7cm]{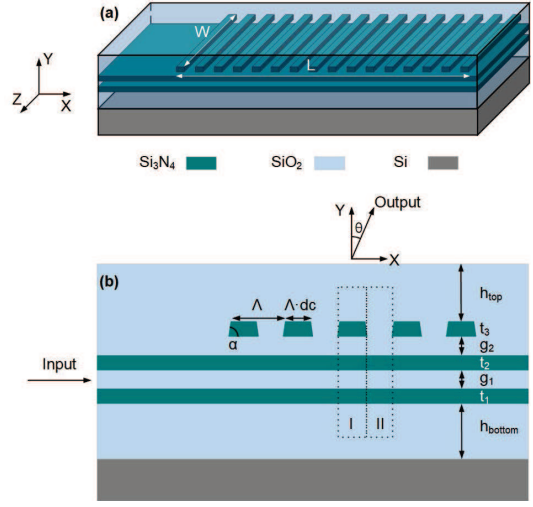}
\caption{Schematic diagram of the structure of the multilayer silicon nitride waveguide diffraction grating. 
(a) 3D view of the grating structure; 
(b) cross-sectional view in the X-Y plane.}
\end{figure}

\subsection{Optimization of the upward efficiency}

The grating diffraction angle $\theta$ in air medium is determined by:

\begin{equation} \label{eq1}	
	{\rm sin\theta}=n_{g}-\frac{\lambda{_{0}}}{\Lambda}
\end{equation}

where $\Lambda$ is the grating period, $n_{g}=(1-dc)\cdot n_{etch}+dc\cdot n_{unetch}$ is the effective index of the waveguide grating, $dc$ is the duty cycle of the grating, and $n_{unetch}$ and $n_{etch}$ are the effective indices of the nonetched and etched waveguides, respectively (see regions I and II in Fig. 1(b)).

To enhance the upward efficiency at a specific diffraction angle, the composite core layers (including the first $\rm Si_{3}N_{4}$ and $\rm SiO_{2}$ layers and the second $\rm Si_{3}N_{4}$ layer) are optimized based on photonic band-gap theory. 
The composite core can be seen as a one-dimensional photonic crystal (1DPC) comprised by $\rm Si_{3}N_{4}$ and $\rm SiO_{2}$ alternating layers, as shown in Fig.2 (a). 
When an incident plane wave falls into the band-gap of a 1DPC, the wave can not propagate in the structure and would be totally reflected. 
This reflection condition can be determined by combining the free space dispersion relation of the incident wave and the off-axis band structure of the 1DPC. 

In the off-axis band structure, the normalized frequency ( $\Omega=\omega a /2\pi c$) of each energy level is a function of the normalized wave vector $K_{x,y}=k_{x,y}a/2\pi$. 
Here,  $\omega$ is the angular frequency of light, a is the period of the 1D photonic crystal, and c is the speed of light in vacuum. 
Fig.2(b) shows the normalized frequency of the first ($n=1$, solid line) and the second ($n=2$, dashed line) energy level versus wave vector $K_{x}$. And for each energy level, only the lower and upper boundary corresponding to $K_{y}=0$ (blue) and  $K_{y}=0.5 $ (red) is shown. When the wave vector is between the center and boundary ($K_{y}\in (0,0.5)$), the dispersion relation curve lies in the region between the blue and red lines. 

In Fig.2(b), the incident wave is TE polarized, and the optical paths of each layer are equal to each other for the normal incident condition; i.e., $t_{1,2}\cdot n_{\rm Si_{3}N_{4}}=g_{1}\cdot n_{\rm SiO_{2}}$ , where $n_{\rm Si_{3}N_{4}}=1.9965$, and $ n_{\rm SiO_{2}}=1.4503$.
And the calculation is performed using Lumerical software based on 2D finite difference time domain (FDTD) algorithm.

Consequently, the overall off-axis band-gap is the gap between the first and the second energy level,  i.e., the section between the dispersion of  $K_{y}=0.5, n=1$ (red solid) and $K_{y}=0.5, n=1$ (red dashed). 

While the dispersion relation of an incident wave with tilt angle $\theta$  in free space is a straight line described by $K_{x}c/\Omega={\rm sin}\theta$. And the intersection of incident wave dispersion relation and the off-axis band-gap is the reflection condition.

For example, a plane wave with an incident angle $\theta=\pm 10^{\circ}$ in air, that is, $K_{x}c/\Omega={\rm sin}10^{\circ}$, is depicted in Fig.2(b) as the solid black line, which intersects with blue and red lines at points A, B, C and O.  
So the dispersion relation on segment BC is in the band-gap, and plane waves incident from $\theta=\pm 10^{\circ}$ in air should be totally reflected. 
Then, the thickness of the composite core layer is determined according to the center point of line segment BC and the equal optical path condition. 
For the wavelength of 850 nm, the photonic crystal period is $a = 255$ nm by the coordinates of points B and C ($K_{x}=0.0674,\Omega=0.3306$) and ($K_{x}=0.0594,\Omega=0.2688$) in Fig. 2. 
The structure coefficients of the grating are obtained by the equal optical path condition, i.e., the thickness of the first and second $\rm Si_{3}N_{4}$ layers $ t_{1}=t_{2}=107nm$, and the thickness of the first $\rm SiO_{2}$ layer $g_{1}=148nm$.

\begin{figure}[htbp]
\centering\includegraphics[width=12cm]{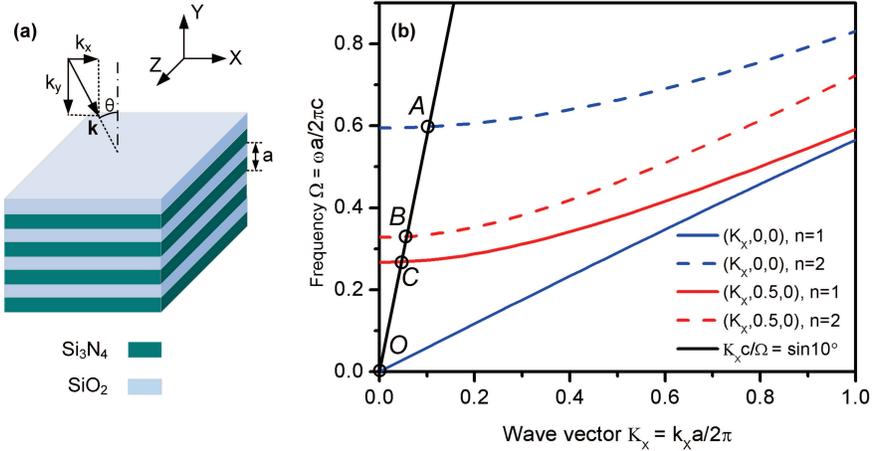}
\caption{Geometry (a) and TE polarized off-axis photonic band structure (b) of a 1D photonic crystal comprising $\rm Si_{3}N_{4}$ and $\rm SiO_{2}$ alternating layers, where $\Omega=\omega a/2\pi c$ is the normalized frequency, $K_{x,y}=k_{x,y} a/2\pi$ is the normalized wave vector, $\omega$ is the angular frequency of light, a is the period of the 1D photonic crystal, and c is the speed of light in vacuum. 
The optical paths of each layer are equal to each other, i.e., $t_{1,2}\cdot n_{\rm Si_{3}N_{4}}=g_{1} \cdot n_{\rm SiO_{2}}$. The solid and dashed blue lines are the dispersion relation of the first and second energy levels, respectively, corresponding to the center of the Brillouin zone ($K_{y}=0$,$K_{z}=0$). 
The solid and dashed red lines are the dispersion relation of the first and second energy levels, respectively, corresponding to the boundary of the Brillouin zone ($K_{y}=0.5$,$K_{z}=0$). 
The solid black line represents a plane wave with incident angle $\theta=10^{\circ}$ in air, that is, $K_{x} c/\Omega={\rm sin}10^{\circ}$.}
\end{figure}

However, the thickness of the composite core in the diffraction grating is limited by the single mode condition. And part of the diffractive beam still passes down through the composite core, and is then reflected on both sides of the lower cladding, leading to a Fabry-Perot interference effect. 
Therefore, the thickness of the bottom cladding $h_{bottom}$ should be set to form constructive interference in the upward direction. The same is true for the choice of the thickness of the top cladding $h_{up}$ and the second spacer layer $g_{2}$.

\subsection{Optimization of the near-field effective length}

The near-field effective length of grating is usually defined by the diffraction length where diffraction intensity decays to $e^{-2}$ of the initial value. And for a uniform diffraction grating, near-field effective length can be determined by 
by $L=1/\alpha_{0} $, where  $\alpha_{0}$ is the grating strength, defined as\cite{4}:

\begin{equation} \label{eq2}	
	2\alpha_0= -\frac{{\rm d}P}{{\rm d}x}\cdot\frac{1}{P}\
\end{equation}
here $P$ is the power in the grating waveguide. 

For the designed grating, the grating strength can be controlled by tuning the etched layer thickness $t_{3}$ and the duty cycle $dc$. 
And, to maintain a high upward efficiency, the diffraction angle should be fixed at $\theta=\pm 10^{\circ}$; consequently, the grating period $\Lambda$ should be adjusted accordingly due to the variation in the effective index (Eq.(1)). 
Also, the upward constructive interference condition should be fulfilled, 
so the distance from the center of the etched layer to the bottom of the second spacer and the top cladding should be constant.

Fig. 3 shows that the upward diffraction efficiency changes with the duty cycle $dc$ and the etched layer thickness  $t_{3}$ at incident wavelength $\lambda_{0}$=850 nm and diffraction angle $\theta= -10^{\circ}$ (a negative angle is chosen to avoid higher-order diffraction). 
When the duty cycle is in the range of 0.1-0.8 and the thickness of the etched layer is in the range of 10-95 nm, the upward diffraction efficiency is always higher than 90$\%$, and the effective length can be adjusted from 100 $\mu m$ to 80,000 $\mu m$.
The result in Fig.3 is simulated by modeling a grating segment of 10 period based on 2D FDTD, which is consistent with the 3D FDTD result but consume less computing resources.

\begin{figure}[htbp]
\centering\includegraphics[width=12cm]{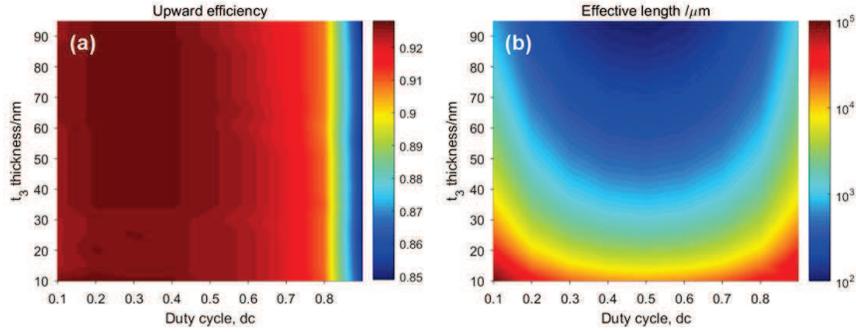}
\caption{Upward diffraction efficiency mapping (a) and effective length mapping (b) versus different etching layer thicknesses and duty cycles in the grating. 
Here, the wavelength $\lambda_{0}$=850 nm, the diffraction angle $\theta= -10^{\circ}$, the thickness of the multilayer $t_{1}=t_{2}$=107 nm,$g_{1}$=148 nm, the thickness of the second oxide spacer layer $g_{2}$=145 nm-$t_{3}$/2, the thickness of the top cladding layer $h_{top}$=3.095 $\mu m$-$t_{3}$/2, and the thickness of the bottom cladding layer $h_{bottom}$=3.390 $\mu m$.}
\end{figure}

\subsection{Simulation of a complete grating}
A diffraction grating with a far-field divergence angle below 0.2$^{\circ}$ at a wavelength of 850 nm is designed and simulated. 
First of all, 376 $\mu m$ is chosen as the effective length, which corresponds to $dc$=0.5 and $t_{3}$=50 nm from Fig. 3. 
Next, the thickness of layers $t_1$=$t_2$=107 nm, $g_1$=148 nm are determined by the reflection condition at $\theta=-10^{\circ}$ , and $h_{bottom}$=3.390 $\mu m$,  $h_{top}$=3.070 $\mu m$, $g_2$=120 nm are obtained by the upward constructive interference condition mentioned in section 2.1. 
Then, the grating period $\Lambda$=464 nm is determined by substituting $\theta$ = -10$^{\circ}, n_{unetch}$=1.6630, $n_{etch}$=1.6503 into Eq.(1). 
Finally, to provide sufficient length for diffraction, the period number is set as 2000, and the total length of the grating is 928 $\mu m$.

The 1D diffraction grating is simulated by the 2D FDTD algorithm and the performance is evaluated. The upward efficiency is obtained by integrating the near-field upward power divided by the input power, the near-field effective length is calculated by fitting the near-field power profile,  and the far-field profile is calculated by Fourier transformation of near-field electric field amplitude profile.

The peak upward diffraction efficiency is 92$\%$ (-0.362 dB) at a wavelength of 850 nm, and the 1 dB bandwidth is 39.8 nm (Fig. 4(a)). 
The residual transmittance is 0.85$\%$, and the reflectance is -36.3 dB. The near-field diffraction intensity decays exponentially as predicted by Eq.(3), and the effective length is calculated to be 376 $\mu m$ (Fig. 4(b)). 
The far-field peak is located at -10.08$^{\circ}$, and the divergence within the full width of $e^{-2}$ is 0.105$^{\circ}$ (Fig. 4(c)).


\begin{figure}[htbp]
\centering\includegraphics[width=12cm]{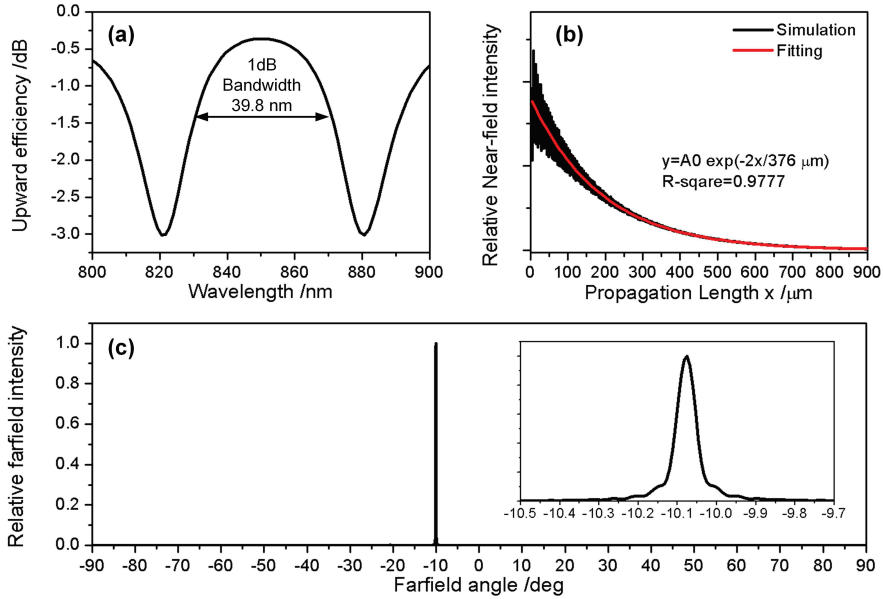}
\caption{Calculated data of the designed 1D diffraction grating. (a) The upward diffraction efficiency versus the incident wavelength; 
(b) the near-field diffraction intensity versus the propagation length in the transmission direction;
(c) the far-field diffraction intensity versus the far-field angle. The inset is a magnified image.}
\end{figure}

\section{Tolerance analysis}
Further, the influence of process variation and laser wavelength drift on the upward efficiency and effective length of the grating are analyzed using the Monte Carlo method. 
Conventional processing deviations, including thickness deviation during CVD deposition, lithographic deviation, and depth deviation during ICP etching, need to be considered and are shown in Table 1.

\begin{table}[htbp]
\centering
\begin{scriptsize}
\caption{\label{tab1} Parameters, processing deviation in the grating and wavelength drift}
\begin{tabular}{ccc}\hline

Parameters &Standard values &Variation \\	
\noalign{\smallskip} \hline	
Thickness of $\rm SiO_{2}$ bottom cladding $h_{bottom}$& 3.39 $\mu m$ & Gaussian, $3\sigma \pm 5\%$\\	
Thickness of first $\rm Si_{3}N_{4}$ layer $t_{1}$& 107 nm & Gaussian, $3\sigma \pm 5\%$  \\	
Thickness of first $\rm SiO_{2}$ layer $g_{1}$& 148 nm & Gaussian, $3\sigma \pm 5\%$  \\	
Thickness of second $\rm Si_{3}N_{4}$ layer $t_{2}$& 107 nm & Gaussian, $3\sigma \pm 5\%$  \\	
Thickness of second $\rm SiO_{2}$ layer $g_{2}$& 120 nm & Gaussian, $3\sigma \pm 5\%$  \\	
Thickness of third $\rm Si_{3}N_{4}$ layer $t_{3}$& 50 nm & Gaussian, $3\sigma \pm 5\%$  \\	
Thickness of top cladding $\rm SiO_{2}$ $h_{top}$& 3.07  $\mu m$ & Gaussian, $3\sigma \pm 5\%$\\	
Grating etching tooth width & 232 nm & Gaussian, $3\sigma \pm$ 50 nm\\	
Grating etching depth & 70 nm & Gaussian, $3\sigma \pm $5 nm\\	
Wavelength & 850 nm & Uniform, $\pm$ 20 nm\\
\hline
\end{tabular}
\end{scriptsize}
\end{table}

Statistical analysis based on the calculation of 800 samples showed that more than 71$\%$ of the samples have an upward diffraction efficiency above 0.8, and $100\%$ of the samples have a nearfield effective length above 200 $\mu m$ (corresponding to a farfield divergence angle of 0.2$^{\circ}$), as shown in Fig. 5.

\begin{figure}[htbp]
\centering\includegraphics[width=12cm]{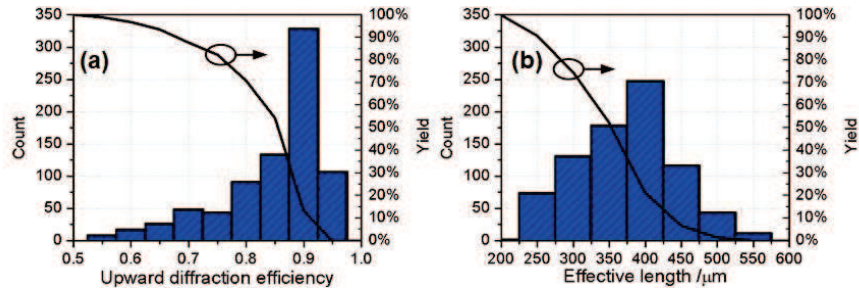}
\caption{Statistical analysis of diffraction grating performance influenced by process variation and incident wavelength drift.
(a) The counts and yield of upward diffraction efficiency; 
(b) the counts and yield of effective length.}
\end{figure}

\section{Discussion}
The proposed slab diffraction grating achieves an upward efficiency of 92$\%$ by optimizing the layers of composite core and cladding, which is much simpler in fabrication than introducing reflector in the cladding of grating.  
Compared with the reported 1550 nm double-layer grating\cite{7}, the grating in this paper has a similar upward efficiency but needs only one step of lithography in the preparation on the diffractive area. Thus the fabrication difficulty is significantly reduced. 
And, compared with the reported 905 nm grating\cite{12}, the far-field divergence angle in this paper is 0.1$^{\circ}$, much less than the reported 0.7$^{\circ}$.
In addition, the effective length of diffraction grating in this paper can be adjusted by changing the thickness of the etching layer. Larger effective length or smaller far-field divergence angle will not increase the difficulty of etching process, because if the minimum etched depth is larger than the thickness of the third etched layer, i.e., over-etching into the second silicon oxide spacer, the influence caused by etching variations on the grating performance is negligible since the over-etched groove will be refilled by silicon oxide after the top cladding is deposited. Therefore, the proposed structure is insensitive to the etching depth errors.
For OPAs with slab grating antenna, crosstalk only occurs among the waveguide arrays connected to the grating, leading to smaller output spacing and larger scanning angle range. Notably, the performance of gratings under different scanning angles is not discussed in this paper and deserves further study.

\section{Conclusion}

A novel diffraction grating based on a multilayer silicon nitride waveguide is designed, and it theoretically achieves an upward diffraction efficiency of 92$\%$ (-0.36 dB), a near-field effective length of 376 $\mu m$ and a far-field divergence angle of 0.105$^{\circ}$ at a wavelength of 850 nm. 
Monte Carlo analysis shows that the designed diffraction grating has a high tolerance for process deviations. 
Under the conditions of $\pm 5\%$ layer thickness variation, $\pm$50 nm lithographic variation and $\pm$20 nm wavelength drift, more than 71$\%$ of the grating samples have a diffraction efficiency higher than 80$\%$ (-0.97 dB), and 100$\%$ of the samples have an effective length larger than 200 $\mu m$ (corresponding to a far-field divergence angle <0.2$^{\circ}$).
Furthermore, by changing the etching layer thickness and the grating duty cycle, the near-field effective length of the grating can be adjusted from hundreds of microns to centimeters, while the upward diffraction efficiency does not decrease seriously. 
The designed diffraction grating has potential application in the region of optical sensing and imaging from visible to near infrared wavelengths.

\section{Funding}
This work was supported by the National Natural Science Foundation of China (NSFC) (No. 62175078 and 61775094).



\end{document}